\documentclass[11pt]{article}
\usepackage{epsfig,psfig,float,amssymb,latexsym}
\usepackage[notref,notcite]{}

\newcommand{\be}{\begin{equation}}
\newcommand{\ee}{\end{equation}}
\newcommand{\ba}{\begin{eqnarray}}
\newcommand{\ea}{\end{eqnarray}}

\newcommand{\nn}{\nonumber}

\begin{document}

\begin{center}
{\Huge{\bf{The rho meson in nuclear matter -\\
\vspace{0.4cm}
a chiral unitary approach 
\footnote{Presented at Mesons and Light Nuclei '01, Prague,
Czech Republic, July 2-6, 2001}}}}
\end{center}
\vspace{.5cm}

\begin{center}
{\huge{D. Cabrera}}
\end{center}

\begin{center}
{\small{\it Departamento de F\'{\i}sica Te\'orica and IFIC,\\
Centro Mixto Universidad de Valencia-CSIC,\\
Institutos de Investigaci\'on de Paterna, Apdo. correos 22085,\\
46071, Valencia, Spain}}
\end{center}

\vspace{1cm}

\begin{abstract}
{\small{In this work, the properties of the $\rho$ meson at rest in cold
symmetric nuclear matter are studied. We make use of a chiral unitary approach 
to pion-pion scattering in the vector-isovector channel, calculated from the 
lowest order Chiral Perturbation Theory ($\chi PT$) lagrangian including 
explicit resonance fields. Low energy chiral constraints are considered by 
matching our expressions to those of one loop $\chi PT$. To account for the 
medium corrections, the $\rho$ couples to $\pi\pi$ pairs which are properly 
renormalized in the nuclear medium, accounting for both $p-h$ and $\Delta -h$ 
excitations. The terms where the $\rho$ couples directly to the hadrons in the
$p-h$ or $\Delta-h$ excitations are also accounted for. In addition, the $\rho$
is also allowed to couple to $N^{*}(1520)-h$ components.

\vspace{0.5cm}

}}
\end{abstract}

\section{Meson-meson scattering in a chiral unitary approach}
We study the $\rho$ propagation properties by obtaining the 
$\pi \pi \to \pi \pi$  scattering amplitude in the $(I,J)=(1,1)$ channel. The
model for meson-meson scattering in vacuum is fully explained in
\cite{Oller:2001ug,PosterJuan}.
We start from the $(I=1)$ $\pi\pi$, $K \bar{K}$ states in the isospin basis:
\ba
\label{isospin}
|\pi\pi\rangle&=&\frac{1}{2}|\pi^+\pi^- - \pi^- \pi^+ \rangle \nn \\
|K\bar{K}\rangle&=& \frac{1}{\sqrt{2}} |K^+K^- - K^0\bar{K}^0 \rangle.
\ea
Tree level amplitudes are obtained from the lowest order $\chi PT$ lagrangians
\cite{gasser1} including explicit resonance fields \cite{gasser2}. We collect
these amplitudes in a $2\times 2$ symmetric $K$ matrix and work in a coupled
channel approach.

The final expression of the $T$ matrix is obtained by unitarizing the tree level 
scattering amplitudes. To this end we follow the N/D 
method, which was adapted to the context of chiral theory in ref. \cite{N/D}. We 
get
\be
\label{T}
T(s)= \left[I+K(s)\cdot G(s) \right]^{-1}\cdot K(s),
\ee
where $G(s)$ is a diagonal matrix given by the loop integral of two meson
propagators. In dimensional regularization its diagonal elements are given by
\be
\label{g(s)}
G_i^D(s)=\frac{1}{16\,\pi^2}\left[-2+d_i+\sigma_{i}(s)\,
\log \frac{\sigma_{i}(s)+1}{\sigma_{i}(s)-1} \right],
\ee
where the subindex $i$ refers to the corresponding two meson state (1 for $K
\bar{K}$, 2 for $\pi\pi$) and 
$\sigma_{i}(s)=\sqrt{1-4 m_i^2/s}$ with $m_i$ the mass of the particles in the
state $i$. The $d_i$ constants in eq. (\ref{g(s)}) are chosen to obey the low
energy chiral constraints,
and they are obtained by a matching to one loop $\chi PT$.

At this stage, the model successfully describes $\pi\pi$ 
P-wave phase shifts and $\pi$, $K$ electromagnetic vector form factors up to 
$\sqrt{s}\lesssim 1.2$ GeV.
The results for the $\pi\pi\to\pi\pi$ scattering amplitude show that the
inclusion of the $K \bar{K}$ channel introduces minimum 
changes compared to the calculation including only pion loops. Keeping this in
mind, the calculations in nuclear matter are performed ignoring the contribution
of kaon loops. 

By using dimensional regularization it is possible to establish connection with
other approaches where tadpole terms are explicitly kept in the lagrangian
\cite{Urban:1998eg}. One can prove that the formalism keeping tadpoles and full off
shell dependence of the $\rho \pi \pi$ vertex is equivalent to the one we have
followed where only the on shell part of the $\rho \pi \pi$ vertex is kept.
In the medium, however, the pion propagator in the tadpole term will change.
Hence, in order to stick to the gauge invariance of the vector field formalism
the tadpole term is kept.

\section{$(I,J)=(1,1)$ $\pi \pi$ scattering in the nuclear medium}

The basic input for the calculation in nuclear matter is
the pion selfenergy. It is written as usual in terms of the Lindhard
functions accounting for both $p-h$ and $\Delta -h$ excitations. Short range
correlations are also accounted for with the Landau-Migdal parameter $g'$,
set to $0.7$. The final expression is
\begin{eqnarray}
\label{Piself1}
\Pi_{\pi}(q,\rho) = f(\vec{q}\,^2)^2 \vec{q}\,^2 \frac{(\frac{D+F}{2 f})^2 \,
U(q,\rho)}{1-(\frac{D+F}{2 f})^2 \, g' U(q,\rho)}, 
\end{eqnarray}
where here $q$ is the four-momentum of the pion and $U=U_N + U_{\Delta}$ the
Lindhard function for $p-h$, $\Delta -h$ excitations
\cite{ReportdeEulogio,Chiang:1998di}. We use a 
monopole form factor $f(\vec{q}\,^2)=\frac{\Lambda^2}{\Lambda^2 + \vec{q}\,^2}$
for the $\pi N N$ and $\pi N \Delta$ 
vertices with the cut-off parameter set to $\Lambda=1$ GeV.
\begin{figure}[ht]
\centerline{\includegraphics[width=0.45\textwidth]{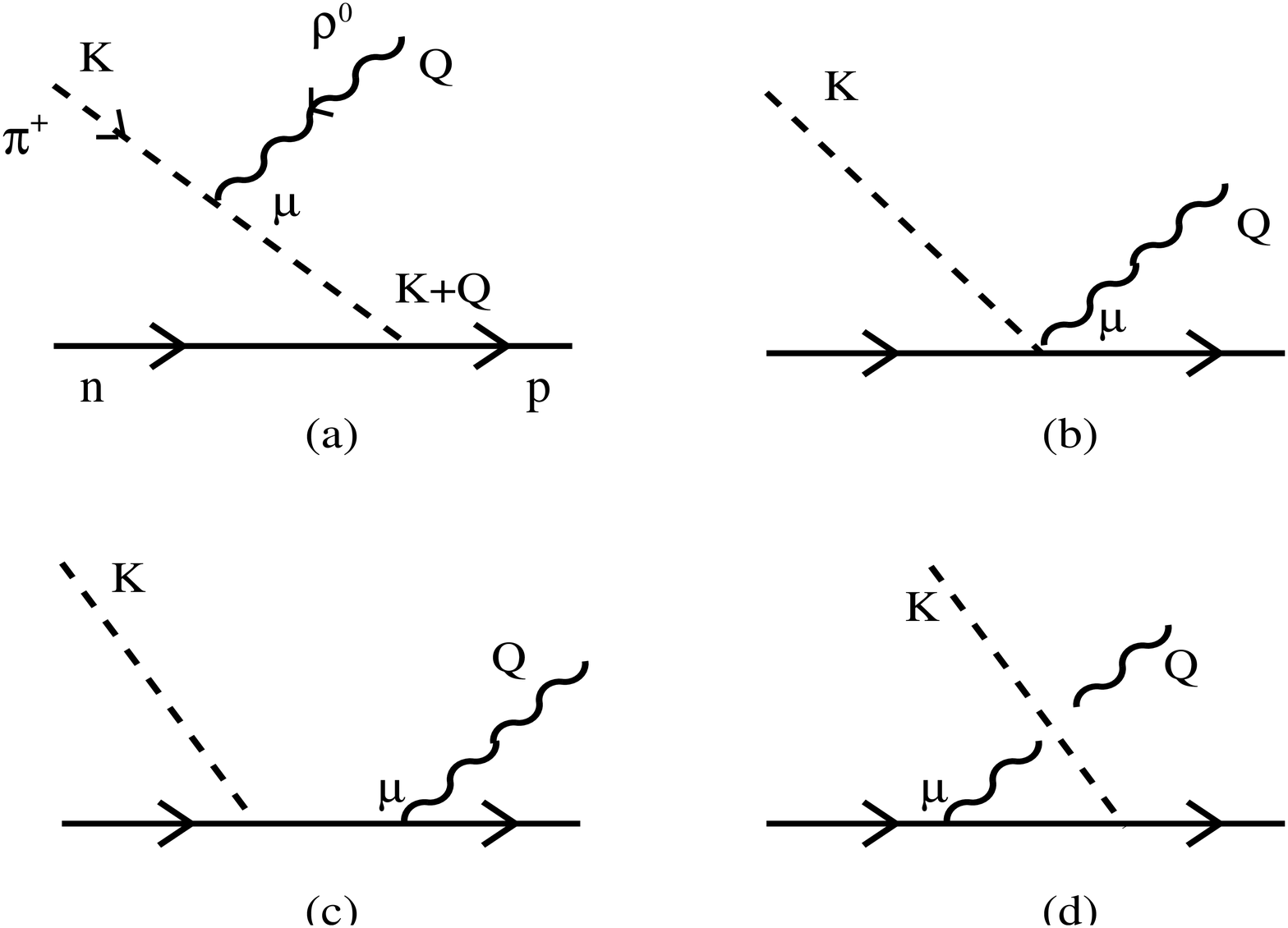}
\hspace{1cm}
\includegraphics[width=0.45\textwidth]{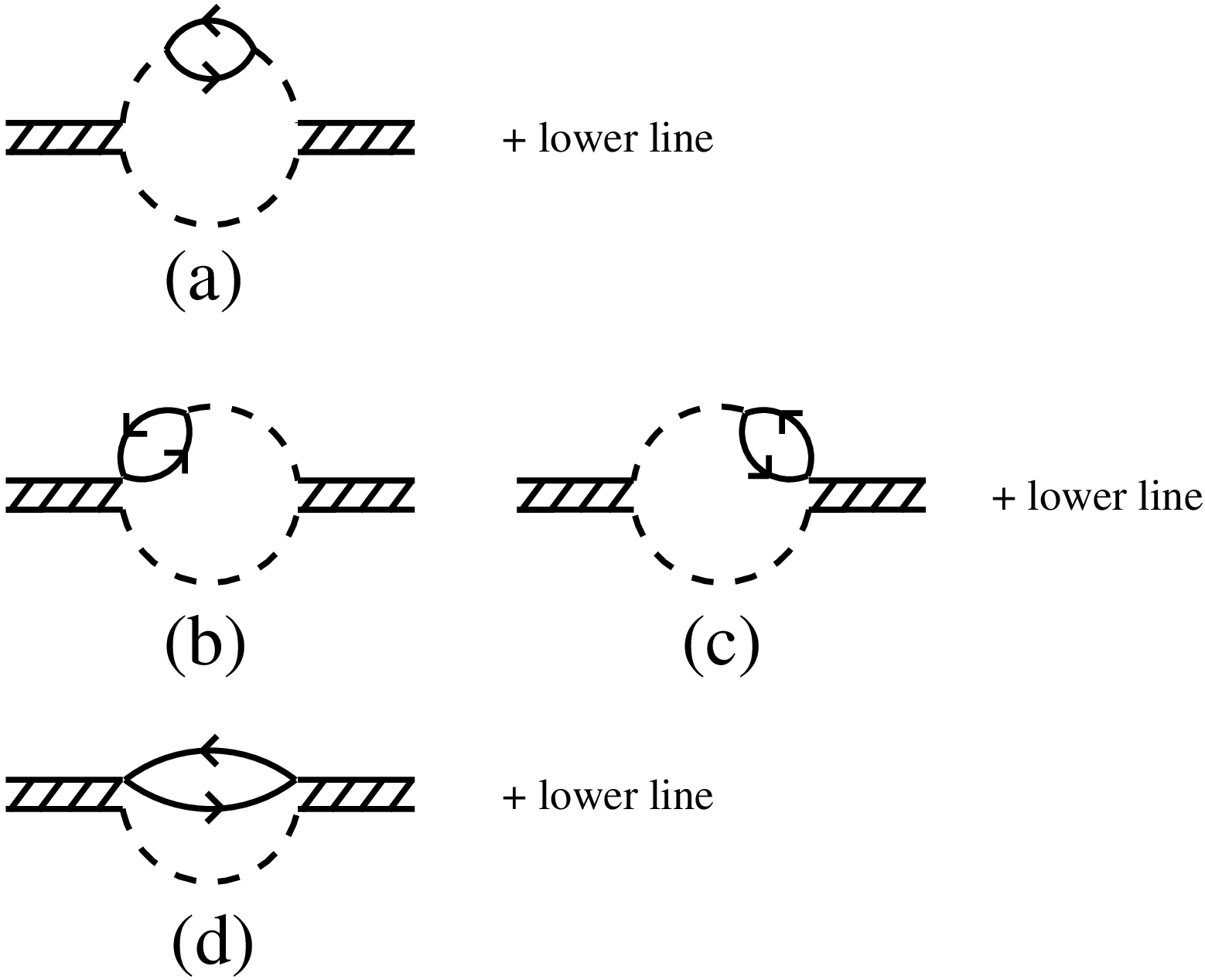}}
\caption{\footnotesize{Left: Gauge invariant set of diagrams used to calculate the 
$\rho$-meson-baryon contact term. The wavy lines represent here the vector
meson, single solid lines are reserved for particle-hole excitations and dashed
lines are $\pi$ mesons. Right: Medium correction graphs. Double solid line
filled with short oblique dashes represents the $K_{22}$ scattering amplitude,
external particle lines are omitted.}}
\label{diagramas}
\end{figure}

In accordance to the gauge invariance of the theory, a $\rho$-meson-baryon
contact term must be considered
\cite{Klingl:1997kf,Chanfray:1993ue,Herrmann:1993za}. Its contribution can be
derived from the set of 
gauge invariant diagrams in Fig. \ref{diagramas} (left), and together with the single
insertion of a pion selfenergy, it provides the set of medium correction graphs
shown in Fig. \ref{diagramas} (right) \cite{Cabrera:2000ct,Cabrera:2000dx}.
Their contribution is readily incorporated in the 
$T$ matrix by a proper substitution and redefinition of the $G$ two meson loop
function of eq. (\ref{T}), and using fully dressed pion propagators. A
subtraction of the contribution in vacuum is performed to cancel quadratic
divergences and convergence is achieved by means of the form factors.

In the same fashion the tadpole diagram in nuclear matter is calculated,
and by subtracting the contribution in vacuum quadratic divergences are
removed. 

\begin{figure}[ht]
\centerline{\includegraphics[width=0.3\textwidth]{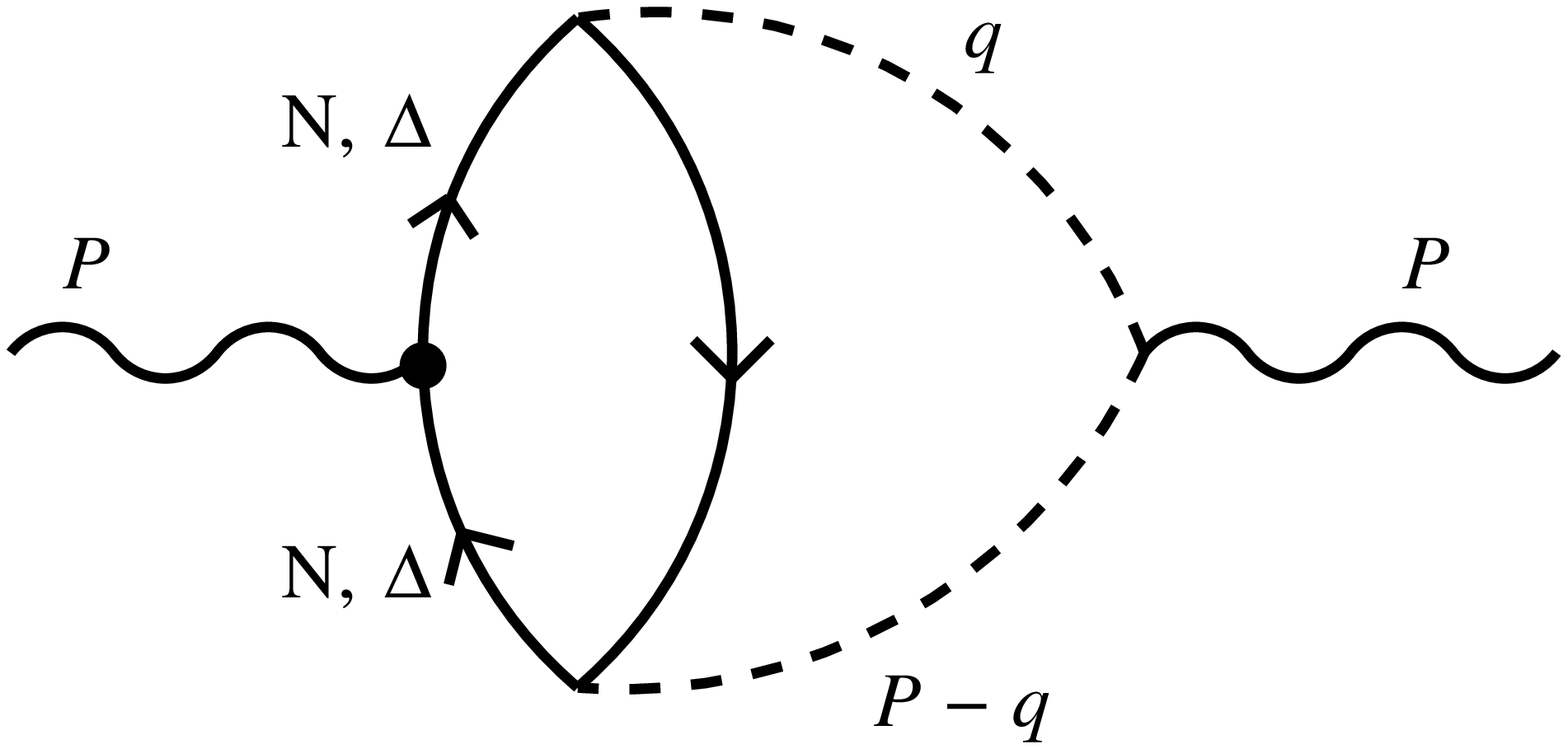}
\hspace{0.5cm}
\includegraphics[width=0.65\textwidth]{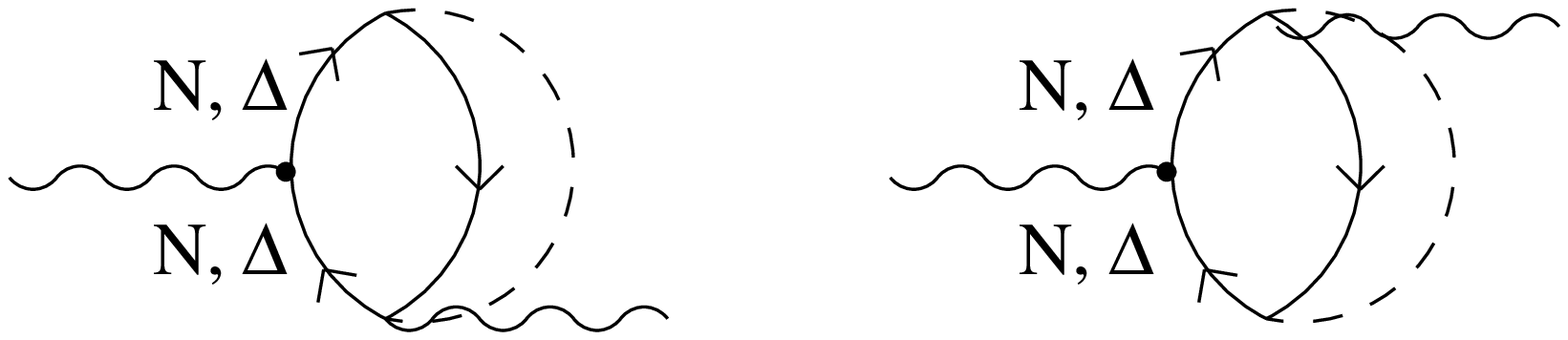}}
\caption{\footnotesize{Additional medium corrections involving $\rho N N$, $\rho
\Delta \Delta$ vertices.}}
\label{additional}
\end{figure}

In addition to the contact interactions mentioned above there are
other medium corrections that arise if one sticks to the gauge vector field
formalism and generate interactions via minimal coupling scheme
\cite{Urban:1998eg}. Because of this a set of diagrams involving $\rho N N$ and
$\rho \Delta \Delta$ vertices also contribute to the $\rho$ meson selfenergy in
nuclear matter. The relevant ones are shown in Fig. \ref{additional}.

\begin{figure}[ht]
\centerline{\includegraphics[width=0.5\textwidth]{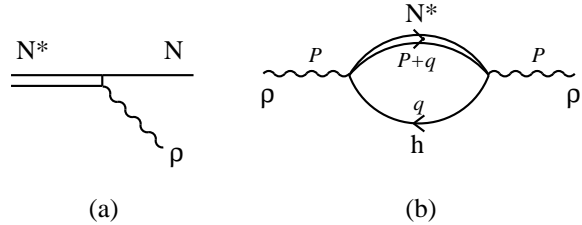}}
\caption{\footnotesize{(a) $\rho N N^*$ vertex. (b) $N^*-h$ bubble contributing
to the selfenergy of the $\rho$ meson.}}
\label{diagNstar}
\end{figure}

A step forward is done by considering the coupling of the $\rho$ meson to
the $N^*(1520)$ resonance. A much detailed work along these lines has been done in
\cite{Peters:1998va}, where many other resonances are included. This correction
manifests as an extra selfenergy term in the $\rho$ propagator. The basic vertex involved in this effect is shown in Fig.
\ref{diagNstar}a, and the lagrangian describing the
interaction reads \cite{osetcanogomez}
\ba
\label{LN*Nrho}
\mathcal{L}_{N^* N \rho} = -g_{N^* N \rho} \bar{\Psi}_N S_i \vec{\phi_i}
\vec{\tau} \Psi_{N^*} + h.c.,
\ea
and the contribution of the selfenergy diagram of Fig. \ref{diagNstar}b can be
written as
\ba
\label{PINstar2}
\Pi_{\rho}^{N^*-h} (P) &=& \frac{2}{3} g_{N^* N \rho}^2 U_{N^*} (P),
\ea
in terms of a Lindhard function for the $N^*-h$ excitation, $U_{N^*} (P)$.

\section{Results and discussion}

We calculated in this approach the real and imaginary parts of $T_{22}$, the
$\pi\pi\to\pi\pi$ scattering amplitude matrix element. The imaginary part shows a
clear broadening of the resonance and the peak position is slightly shifted
upwards by about 30 MeV, which is also observed in the position of the zero
of the real part. The coupling to the $N^* -h$ components manifests as
a visible bump at lower energies. As a whole much strength is spread below the
resonance mass.

In order to test the model dependence on the phenomenological
parameters we performed variations of $\Lambda$ in the range 0.9-1.1 GeV, and of
$g'$ in the range 0.6-0.8. The results are rather stable under the first of the
variations, showing uncertainties in the position of the $\rho$ meson peak of
about 10-15 MeV. Variations of $g'$ are relevant close to the resonance maximum,
and lead to uncertainties in the position of the resonance of about 20-25 MeV.
This was expected 
since the pion selfenergy, which is one of the basic ingredients of the medium
corrections, is strongly modified by the short range correlations which directly
depend on $g'$.

In considering the coupling to baryonic resonances our model does not try to be
complete since many other resonances should be included, but this allows us to
have an estimate of how these new channels affect the results. The calculation
could be improved by following a self-consistent treatment as done in
\cite{Peters:1998va}.

\end{document}